\documentclass[conference]{IEEEtran}
\IEEEoverridecommandlockouts

\usepackage{cite}
\usepackage[hidelinks]{hyperref}
\usepackage{amsmath,amssymb,amsfonts}
\usepackage{algorithmic}
\usepackage{graphicx}
\usepackage{textcomp}
\usepackage{tabularx}
\usepackage{multirow}
\usepackage{makecell}
\usepackage{orcidlink}
\usepackage{booktabs}
\usepackage{array}

\def\BibTeX{{\rm B\kern-.05em{\sc i\kern-.025em b}\kern-.08em
    T\kern-.1667em\lower.7ex\hbox{E}\kern-.125emX}}
\begin{document}

\title{Potential of Quantum Computing Applications for Smart Grid Digital Twins and Future Directions}

\author{
Arianne Ornella Lemo N.\orcidlink{0009-0002-4871-5059}, 
Ahmad Mohammad Saber\orcidlink{0000-0003-3115-2384}, \IEEEmembership{Member, IEEE},  \\
Deepa~Kundur\orcidlink{0000-0001-5999-1847},~\IEEEmembership{Fellow,~IEEE},
and 
Adam W. Skorek\orcidlink{0000-0000-0000-0000}, \IEEEmembership{Fellow, IEEE}
\thanks{Arianne Ornella Lemo N. and Adam W. Skorek are with the ECE Department, University of Quebec at Trois-Rivières, Trois-Rivières, QC, Canada (emails: \href{mailto:arianne.lemo@uqtr.ca}{arianne.lemo@uqtr.ca}, \href{mailto:adam.skorek@uqtr.ca}{adam.skorek@uqtr.ca}). Ahmad Mohammad Saber and Deepa Kundur are with the ECE Department, University of Toronto, Toronto, ON, Canada (emails: \href{mailto:ahmad.m.saber@ieee.org}{ahmad.m.saber@ieee.org}, \href{mailto:dkundur@ece.utoronto.ca}{dkundur@ece.utoronto.ca}).}
}


\maketitle

\begin{abstract}

 The convergence of digital twin technology and quantum computing is opening new horizons for the modeling, control, and optimization of smart grid systems. This paper reviews the current research landscape at the intersection of these fields, with a focus on how quantum algorithms can enhance the performance of digital twins in smart energy systems. We conduct a thematic literature review and identify key research trends, technical challenges, and gaps in real-world adoption. Further, a conceptual framework is proposed to integrate quantum modules into classical digital twin architectures.
 The potential benefits of this hybrid approach for smart grid operation and future research directions are also discussed.

\end{abstract}

\begin{IEEEkeywords}
Smart Grid, Security, Digital Twin, Machine learning.
\end{IEEEkeywords}

\section{Introduction}

In recent years, the energy sector has undergone a profound transformation as a result of the convergence of emerging technologies, including but not limited to digital twins, quantum computing, machine learning and artificial intelligence. Digital twins, which are digital representations of real-world systems that monitor, predict and emulate the behavior of complex infrastructure between physical instances, are receiving considerable momentum in the context of smart cities, smart energy systems, and smart grids. Although digital twins were originally used notably in manufacturing and aerospace, it has now been suggested that they should be similarly integrated into power systems in general to improve the design, operation and maintenance of smart grids.
This integration is timely and essential, as the operation of smart grids is becoming increasingly complex due to the rapid proliferation of distributed energy resources, the growing need for sustainability, and the demand for scalable and adaptive control frameworks. Quantum computing, with its capability to process high-dimensional optimization problems, and digital twins, with their ability to mirror real-time grid states, offer complementary strengths that can jointly address these emerging challenges.
At the same time, quantum computing has also expressed a very early potential to address optimization and simulation problems that would not be tractable for classical computers. The unique quantum capabilities of quantum algorithms ignited a new research agenda when combined with smart grid applications: in particular, load forecasting, optimal-despatch of energy, anomaly detection, and real-time monitoring of the grid can all benefit from the application of quantum computing methods \cite{pr12020270}. Although still in its infancy, the combination of digital twins and quantum methods has significant potential to help develop more resilient, more efficient, and ultimately more sustainable smart energy systems.

This paper presents a short review of
the current research landscape at the interplay of digital twins, quantum computing, and smart grid applications. We discuss use cases and technical challenges and possible future paths of digital twins with quantum computing technologies. The aim is to provide a timely update to inform ongoing and future research and establish a foundation for collaborative research across energy, artificial intelligence researchers, and quantum scientists.
To achieve this, a thematic literature review was conducted on recent peer-reviewed publications from 2022 to 2025, covering topics such as quantum optimization, digital twin simulation, and smart grid applications. The selection process focused on identifying recurring research patterns and technological gaps. The results are categorized into major themes, presented in Section II, and supported by an applied framework proposal in Section III that integrates quantum computing modules into smart grid digital twins. Section V discusses practical implications and identifies research gaps, while Section VI concludes the paper and outlines directions for future work.

%

\section{Related Work and Analysis}
Recent advances in high-performance computing (HPC) infrastructure have played a significant role in enabling quantum simulations and hybrid digital twin architectures. Access to cloud-based quantum processors and emulators, as well as national supercomputing platforms, now allows researchers to experiment with quantum-classical models in realistic energy scenarios. These infrastructures support scalable experimentation with quantum-enhanced optimization, simulation, and learning techniques, which are often computationally intensive.
Beyond smart grids, quantum computing has begun to influence various domains within cyber-physical systems (CPS) \cite{goyal2024integrating}, such as autonomous transportation networks \cite{10734228}, aerospace systems \cite{9843455}, and industrial control systems \cite{engproc2025087068}. In these areas, quantum algorithms have been proposed for real-time decision-making, multi-agent coordination, secure communication, and system reliability modeling. These interdisciplinary applications highlight the broader potential of quantum computing in complex CPS architectures and reinforce its relevance as a transformative technology for critical infrastructure.
To structure the key findings of our literature review, we identified six dominant research themes that represent the convergence between quantum computing and digital twin applications in smart grids. Table~\ref{tab:themes} summarizes these themes, their typical applications, and the number of studies contributing to each. This organization facilitates a clearer understanding of the current research landscape and serves as a reference for subsequent thematic discussion in the following subsections.
\begin{table}[t!]
\centering
\caption{Thematic Classification of Reviewed Studies on Quantum Computing and Digital Twins in Smart Grid Applications}
\label{tab:themes}
\begin{tabular}{p{2.5cm}|p{3cm}|c}
\hline
\textbf{Theme} & \textbf{Applications in Smart Grids} & \textbf{No. Ref} \\ \hline
Quantum-Enhanced Digital Twins & Load forecasting, optimization of dispatch, fault anticipation, power flow balancing & \cite{lemo2025modelisation,1571112244} \\ \hline
Quantum-Inspired ML and Data Intelligence & Time-series prediction, anomaly detection, virtual diagnostics & \cite{10549913} \\ \hline
Federated Learning and Scalability & Distributed digital twins, collaborative model training across grid regions & \cite{mathur2025federatedlearningmeetsquantum} \\ \hline
Hybrid Quantum and Classical Interoperability & API standardization, co-simulation between quantum cloud and local DTs & \cite{demaio2025roadhybridquantumprograms,abdullah2024uncertaintysupplychaindigital} \\ \hline
Cybersecurity and Quantum-Resistant Architectures & QKD, quantum-safe encryption, threat modeling & \cite{10969886,10933564,bishwas2024strategicroadmapquantumresistant} \\ \hline
Ethical, Regulatory, and Quantum Governance & Validation protocols for probabilistic quantum outcomes, compliance in critical infrastructure & \cite{herian2025much} \\ \hline
\end{tabular}
\end{table}
For example, the Quantum Improved Weather Forecast (QWF) framework \cite{10391714} illustrates how quantum machine learning, such as quantum SVMs and neural networks, can significantly enhance prediction accuracy in complex, chaotic systems like weather forecasting, or other examples\cite{Abreu_2024,ozguler2025performanceevaluationvariationalquantum}. Further, Table~\ref{tab:ref_summary} depicts a summary of recent quantum computing applications in different fields.

\begin{table}[ht]
\centering
\caption{Summary of Selected References and Their Applications}
\begin{tabular}{p{0.3cm}|p{1cm}|p{1.5cm}|p{1.1cm}|p{3cm}}
\hline
\textbf{Ref.} & \textbf{Quantum Algorithms Used} & \textbf{Smart Grid Application} & \textbf{Dataset / Test System} & \textbf{Key Findings} \\
\hline
\cite{ goyal2024integrating} & Quantum, AI & cyber-physical systems & Conceptual & Opportunities and challenges combining AI with quantum in CPS. \\
\hline
\cite{10734228} & Quantum federated learning & Voltage stability via DT & Simulated power grid & Enhanced voltage stability in smart grids. \\
\hline
\cite{9843455} & -- & Aerospace Applications & Conceptual &  Benefits of quantum computing for aerospace systems. \\
\hline
\cite{engproc2025087068} & Quantum algorithm & Diesel fuel purification & Chemical simulation &  quantum fuzzy control system for diesel fuel hydrotreatment. \\
\hline
\cite{lemo2025modelisation} & QAOA & MV circuit breakers & Simulated gas data &  Quantum estimation of gas insulation properties. \\
\hline
\cite{1571112244} & QAOA + DL & Digital twin for MV circuit breaker & COMSOL + Qiskit & Demonstrates a hybrid digital twin approach. \\
\hline
\cite{10549913} & QML and GA & Enhancing optimization strategies & Algorithmic comparison & Enhancing genetic algorithm design using quantum methods. \\
\hline
\cite{mathur2025federatedlearningmeetsquantum} & Quantum federated learning & energy networks & Simulated grid scenarios & The impact of quantum computing into federated learning for DT. \\
\hline
\cite{demaio2025roadhybridquantumprograms} &  Hybrid quantum-classical & Quantum-classical implementation & Hybrid testbed  &  Roadmap for hybrid system architecture. \\
\hline
\cite{abdullah2024uncertaintysupplychaindigital} & Hybrid quantum-classical & Financial risk via supply chain DTs & Simulated supply chain & Introduces uncertainty quantification via QML. \\
\hline
\cite{10969886} & Quantum, AI & Cybersecurity in 6G & 6G simulation & Securing 6G networks in the era of AI and quantum computing. \\
\hline
\cite{10933564} & quantum cryptography & IoT Healthcare with DT & Simulated IoT network & Secure DT in IoT healthcare using quantum-resistant cryptography. \\
\hline
\cite{bishwas2024strategicroadmapquantumresistant} & Quantum key distribution & Preventing quantum attacks in industry & Threat simulation & Proposes an approach to prevent quantum attacks in industry at the advent of quantum algorithms. \\
\hline
\cite{10391714} & QML & Enhanced weather prediction & Real-world datasets &  Application of quantum AI to enhance weather forecasting accuracy. \\
\hline
\cite{Abreu_2024} & QML & Detecting cyberattacks & Simulated intrusion dataset & Demonstrates QML’s advantage in anomaly detection. \\
\hline
\cite{ozguler2025performanceevaluationvariationalquantum} & VQE & Simulation of quantum dynamics & Quantum simulation & Evaluates performance of quantum simulation models. \\
\hline
\cite{SAINI202537} & QML and DT & Smart cities & Urban models & Explores synergies of QML with DTs in cities. \\
\hline
\end{tabular}
\label{tab:ref_summary}
\end{table}

\subsection{Quantum-Enhanced Digital Twins for Smart Grids}
\begin{figure}[t!]
    \centering
    \includegraphics[width=1\linewidth]{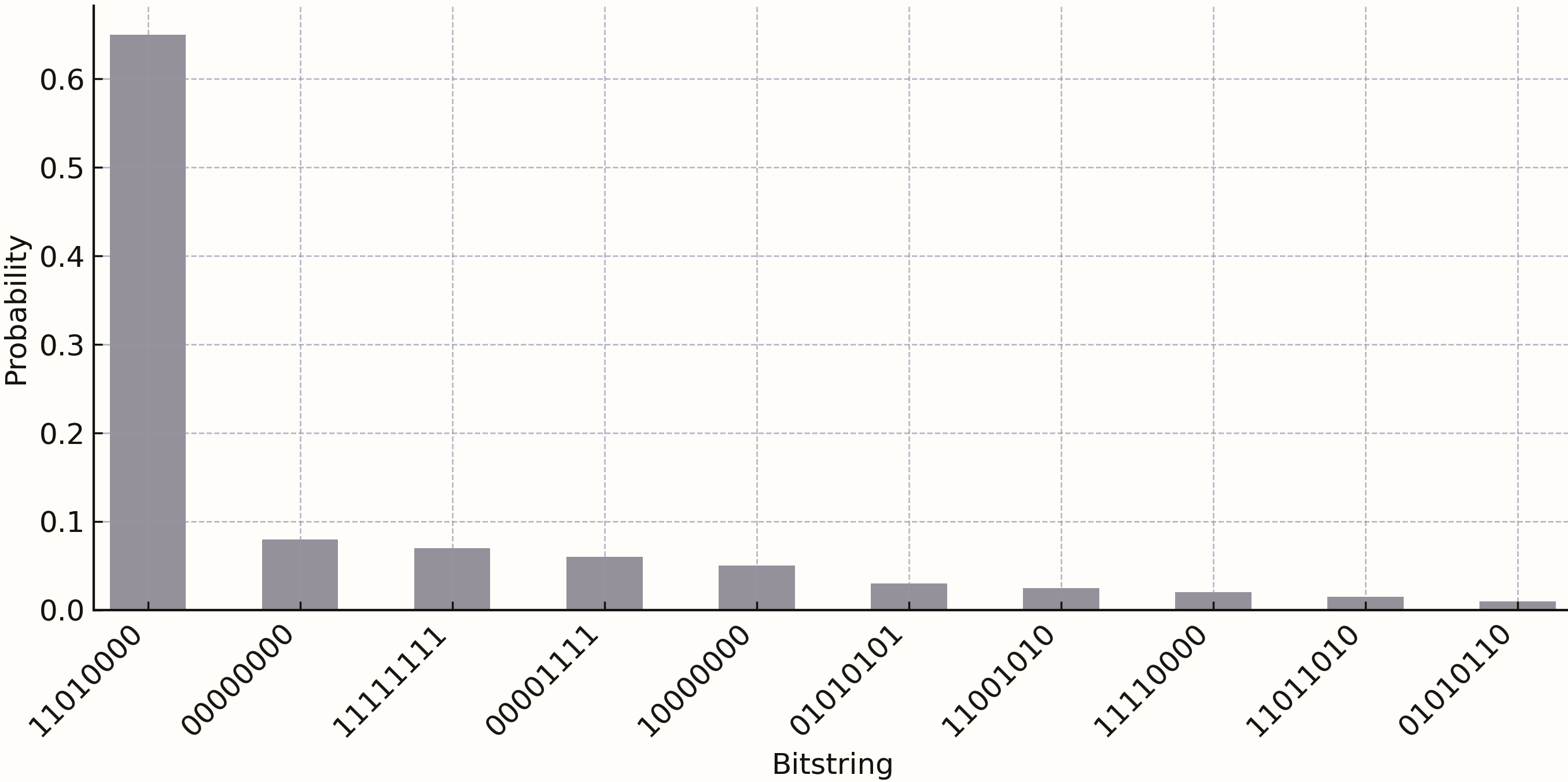}
    \caption{Probability distribution of the bitstrings generated by QAOA \cite{1571112244}.}
    \label{fig:qaoa_result}
\end{figure}

Digital twins (DTs) are moving away from just being simulation tools to being intelligent systems that utilize quantum computing. The use of quantum algorithms in the DT space like the  Quantum Approximate Optimization Algorithm (QAOA) increases DT solution capability with respect to complex optimization problems, such as power flow balance, network topology reconfiguration, and fault anticipation and enables them to do so with greater speed and global accuracy than classical methods. This melding provides real-time resilience and the necessary decision-making capabilities for dynamic smart grids \cite{SAINI202537}.
Table~\ref{tab:comparison_dt} provides a comparative overview between traditional digital twins and quantum-enhanced digital twins. It highlights key differences in terms of scalability, speed, and decision-making capabilities. QEDTs demonstrate superior performance in managing complex smart grid operations by leveraging quantum algorithms for faster and more accurate optimization.
\begin{table}[t!]
\centering
\caption{Comparison between Classical and Quantum-Enhanced Digital Twins}
\label{tab:comparison_dt}
\begin{tabular}{p{4cm} | p{4cm}}
\hline
\textbf{Classical Digital Twin} & \textbf{Quantum-Enhanced Digital Twin} \\ \hline
Limited sequential optimization \cite{yakubova2025application,ali2023quantum} & Parallel optimization using superposition \cite{engproc2025087068} \\ \hline
Relies on traditional HPC resources \cite{iraola2025hp2c,kasztelnik2023digital} & Can leverage hybrid quantum circuits \cite{mathur2025federatedlearningmeetsquantum,abdullah2024uncertaintysupplychaindigital} \\ \hline
Sensitive to dimensionality limitations \cite{10498081} & Better suited for NP-complete problems \cite{bishwas2024strategicroadmapquantumresistant} \\ \hline
Frequently converges to local solutions \cite{ullah2022quantum} & Higher chance of global convergence \cite{ullah2022quantum}\\ \hline
High energy consumption for simulations \cite{jameil2025quantum} & Potential reduction in energy cost (as technology matures) \cite{10549913} \\ \hline
\end{tabular}
\end{table}
Further, recent work in \cite{lemo2025modelisation} designed a digital twin for a medium-voltage circuit breaker using the QAOA. The study reformulated the equations governing dielectric property estimation in gas insulation into a form suitable for quantum execution on both real hardware and simulators. QAOA produced a probability distribution of bitstrings, with the most probable one representing the optimal solution. Fig.~\ref{fig:qaoa_result} highlights this distribution. These results illustrate how quantum algorithms can address complex parameter estimation problems in electrical engineering, such as dielectric strength in gas-insulated systems, and how quantum-enhanced optimization can be integrated into digital twin models for smart grids to improve prediction, adaptation, and efficiency.

\subsection{Quantum-Inspired Data Intelligence and ML Integration}
The vast data created by smart grid digital twins (DTs) benefits from quantum machine learning (QML) for advancing classification, forecasting, and optimization approaches. For example, variational quantum classifiers and quantum-enhanced LSTM models are capable of incorporating recent ways to capture complex and often non-linear correlations in time-series energy data. Full-scale QML implementation is still emerging, but even semi-quantum acceleration via QML could provide training time savings and improved generalization on scarce data \cite{10549913}. This is particularly relevant to emerging grid configurations.

\subsection{Scalability of Digital Twin through Quantum-Assisted Federated Architectures}
The scalability issue of digital twins in national-scale energy systems, referred to as federated learning, can be solved using quantum-assisted federated learning \cite{mathur2025federatedlearningmeetsquantum}. In this scenario, each local DT applies local data, but collectively work for a global optimization objective. Quantum processors will help accelerate the federated optimization or can be used to solve part of the sub-problems, making sure the security and scalability learning through distributed infrastructures.

\subsection{Interoperability in Hybrid Quantum-Classical Environments}
Future smart grid architectures will include heterogeneous computing resources, blending edge devices, classical cloud services, and quantum processors \cite{demaio2025roadhybridquantumprograms}. DTs must adapt to orchestrate workflows across these platforms. Standardized APIs and interoperable modeling layers are needed for DTs to request quantum computation from remote quantum backends and integrate the results seamlessly into real-time decision cycles \cite{abdullah2024uncertaintysupplychaindigital}.

\subsection{Cybersecurity and Quantum-Resistant Architectures}
As Digital Twins (DTs) are at the heart of grid operations, we shift towards a more critical interplay between cyberspace and physical infrastructures. Quantum computing adds value in two areas. Firstly, quantum-enhanced threat modeling provides more real-time and exhaustive simulations of potential attack surfaces; secondly, the development of quantum-resistant algorithms \cite{10969886} for cryptography where data will be exchanged both within and across DTs, especially in the long term. Furthermore, hybrid quantum-classical DTs can also better detect and react to problematic behavior patterns in near real-time \cite{10933564}.  Industries must prepare for the advent of quantum in the world of crytography \cite{bishwas2024strategicroadmapquantumresistant}. 

\subsection{Ethical, Regulatory, and Quantum Governance}
Today, it is useful to consider the regulations governing the use of quantum technology in DTs. How does one validate a result from probabilistic quantum algorithms? What protocols regarding standards and consistency should apply to quantum-assisted decisions in critical infrastructure? These questions put great pressure on policymakers tasked with preparing for the governance of hybrid systems where quantum quantum-assisted models may impact people and energy security. Transparent, clear, and explainable validation protocols can build public trust and promote industrial uptake \cite{herian2025much}.

\section{Illustrative Use Case}

To highlight the practical relevance of integrating quantum computing into digital twins for smart grids, we present a hypothetical yet realistic use case. The scenario involves a microgrid composed of distributed energy resources (DERs), such as photovoltaic panels, battery storage systems, and controllable loads, deployed in a semi-urban residential area.
A digital twin continuously replicates the microgrid’s real-time state, including load profiles, generation capacities, battery state-of-charge, and grid topology. This virtual replica serves for both operational monitoring and decision-making support.
In this context, we propose integrating a quantum optimization module, based on the Quantum Approximate Optimization Algorithm, into the digital twin. The objective is to optimize energy dispatch in real time while considering multiple dynamic constraints such as: minimizing energy loss and cost, maximizing the use of renewable energy, balancing load across nodes, managing battery charging and discharging schedules.

\begin{figure}[t!]
    \centering
    \includegraphics[width=1\linewidth]{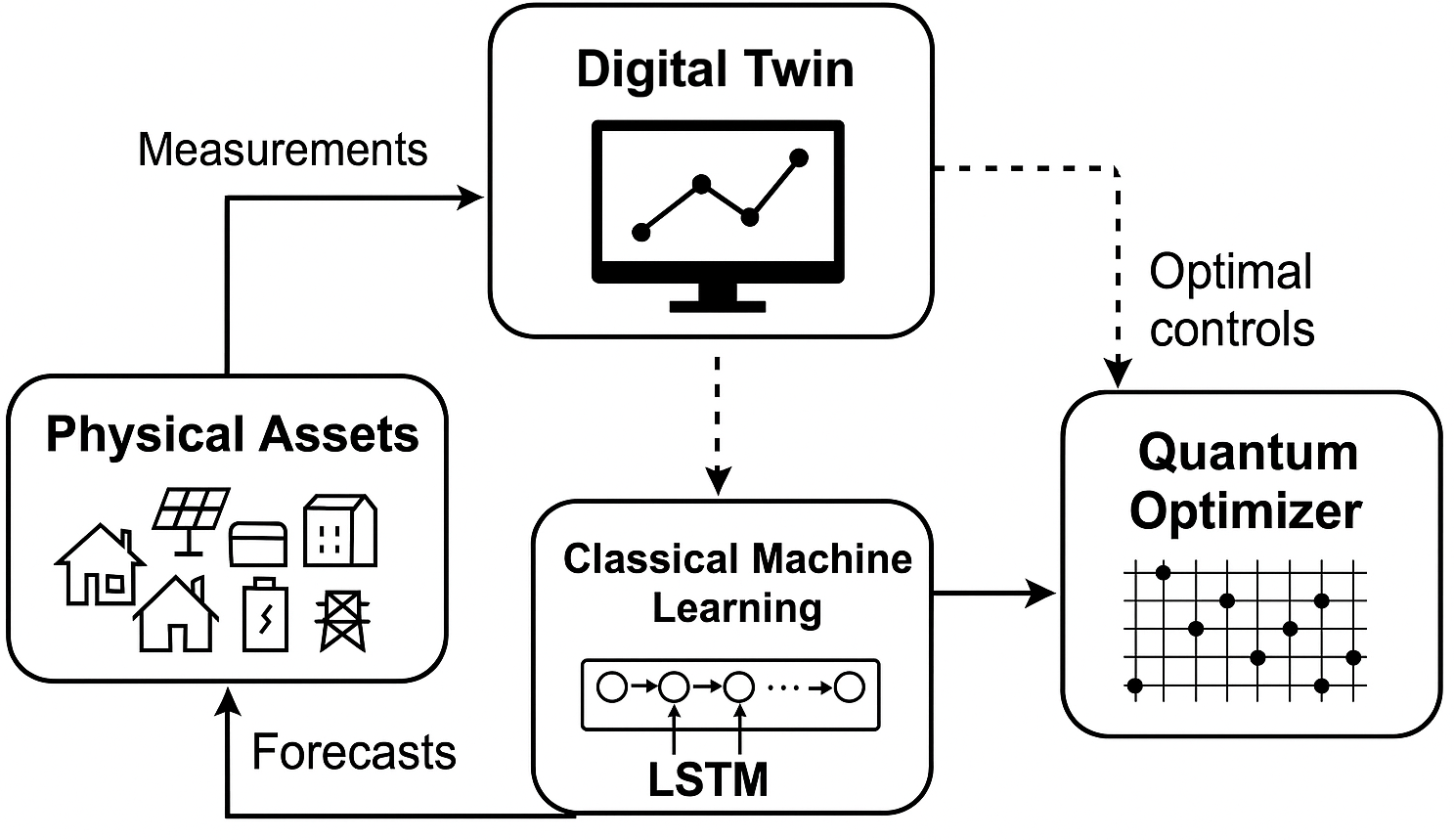}
    \caption{A schematic diagram illustrating a quantum-enhanced digital twin for smart grid systems.}
    \label{fig:quantum_digital_twin}
\end{figure}

While traditional optimizers often struggle with the combinatorial nature of this problem, especially under high variability and uncertainty QAOA leverages quantum superposition and entanglement to explore a broader solution space and converge faster towards global optima.
In parallel, a Long Short-Term Memory (LSTM) network is trained on historical time series data to forecast short-term demand and generation, enabling the digital twin to anticipate system states. The QAOA module then uses these predictions to determine optimal control actions over a given planning horizon.
Figure ~\ref{fig:quantum_digital_twin} depicts the interaction between physical assets, the digital twin, the classical machine learning module (LSTM), and the quantum optimizer. This hybrid architecture exemplifies how quantum-classical co-processing can enhance energy management decisions in future smart grids.

\begin{figure}[t!]
    \centering
    \includegraphics[width=1\linewidth]{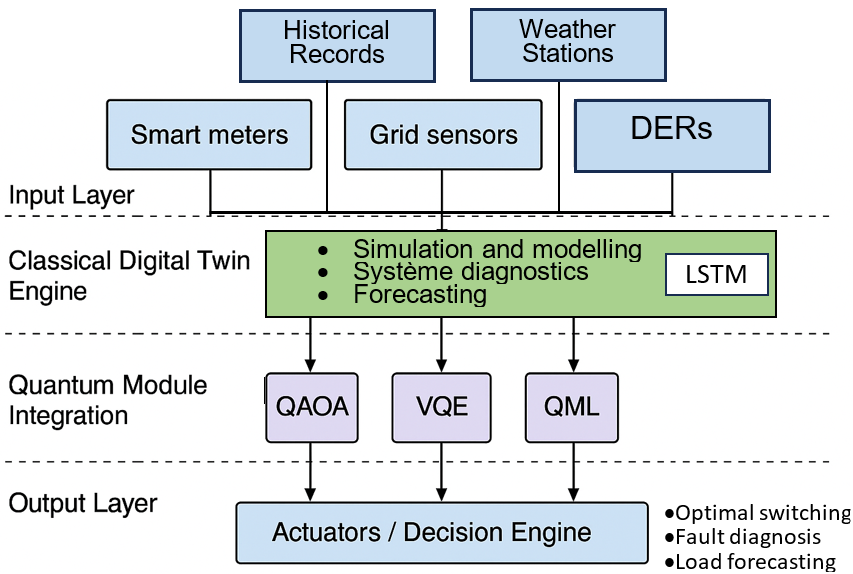}
    \caption{Proposed Hybrid Architecture Combining Quantum Computing and Digital Twins for Smart Grids.}
    \label{fig:quantum_digital_twin_framework}
\end{figure}

This use case highlights the feasibility of combining quantum algorithms with digital twin architectures for smart grid applications, even in the absence of fully matured quantum hardware.
To further guide future research on quantum-enhanced smart grid digital twins, we propose a conceptual framework that integrates classical digital twin architecture with quantum computing modules.
The proposed framework is structured around four main components:
\begin{enumerate}
    \item Input Layer: The system ingests real-time data from smart meters, grid sensors, distributed energy resources (DERs), weather stations, and historical records. Additionally, network topologies and system configuration parameters are included to provide structural context.
    \item Classical Digital Twin Engine: This core engine simulates and replicates the behavior of the physical grid using physics-based and data-driven models. It performs system diagnostics, real-time state estimation, and scenario-based forecasting under normal operation.
    \item Quantum Module Integration: Specific tasks that require high-dimensional optimization—such as predictive maintenance, anomaly detection, or optimal resource dispatch—are offloaded to quantum-enhanced submodules, for instance: (i) Quantum Approximate Optimization Algorithm for solving combinatorial optimization problems (e.g., reconfiguration after faults), (ii) Quantum Machine Learning (QML) for pattern recognition in load demand or cyberattack signatures \cite{Abreu_2024,hammadia2025quantum}, and (iii) Variational Quantum Eigensolver (VQE) for evaluating energy-based models in microgrid dynamics \cite{ozguler2025performanceevaluationvariationalquantum}.
    \item Output Layer: The framework provides actionable insights, such as optimal switching decisions, fault diagnosis, load forecasting, and risk alerts. It supports both operational control and long-term planning in smart grids.
\end{enumerate}
Fig.~\ref{fig:quantum_digital_twin_framework} illustrates this hybrid framework, emphasizing the interface between classical and quantum layers.
This figure presents a high-level schematic of the proposed hybrid architecture, which combines the strengths of quantum computing and digital twin technologies to enhance smart grid performance. The digital twin continuously mirrors the physical smart grid system, enabling real-time monitoring, predictive maintenance, and anomaly detection. Simultaneously, quantum algorithms, specifically quantum optimization or quantum enhanced machine learning, are leveraged to perform complex computations such as forecasting energy demand, optimizing resource allocation, and enhancing cybersecurity defenses, as previously illustrated in
Table~\ref{tab:ref_summary}.

\section{Discussion: Practical Implications}
The comparison between theoretical advancements and industrial realities revealed significant disconnects that must be addressed before quantum-enhanced digital twins can be adopted in smart grid infrastructures. Current limitations include the noise of NISQ devices, the lack of standard interfaces with existing digital twins, and the need for problem-specific quantum encoding strategies, it becomes difficult to validate simulation results or explore realistic deployment scenarios.
Furthermore, the absence of modular and scalable frameworks prevents the seamless integration of quantum modules into existing digital twin systems \cite{10926904} . Most current models are not designed to support hybrid classical-quantum execution, which slows down experimentation and industrial adoption.
Another notable gap is the limited collaboration between quantum computing researchers and energy system engineers. This siloed development leads to innovations that are either too theoretical or misaligned with operational constraints in smart grids. Bridging this gap through interdisciplinary projects could accelerate the creation of usable quantum-enhanced platforms.
Finally, we observed that most quantum applications in energy remain confined to cryptography and algorithm design. There is still a lack of applied use cases, especially in digital twin design, predictive maintenance, and grid optimization. This reinforces the relevance of our study, which contributes to expanding quantum computing into the broader modeling and operation of cyber-physical systems like smart grids.

\section{Conclusion and Future Work}
This study introduced a hybrid quantum-classical approach to support the development of quantum-enhanced digital twins for smart grids. Through simulation results and architectural analysis, we showed the potential of quantum algorithms such as QAOA to address complex optimization tasks in energy systems. Future research directions include developing hybrid architectures combining quantum-based modules with classical DT platforms; validating scalability through implementation on cloud-based quantum processors; and exploring AI-enhanced decision-making loops using quantum-trained models. 

\bibliographystyle{IEEEtran} 

\bibliography{References}

\begin{thebibliography}{10}
\providecommand{\url}[1]{#1}
\csname url@samestyle\endcsname
\providecommand{\newblock}{\relax}
\providecommand{\bibinfo}[2]{#2}
\providecommand{\BIBentrySTDinterwordspacing}{\spaceskip=0pt\relax}
\providecommand{\BIBentryALTinterwordstretchfactor}{4}
\providecommand{\BIBentryALTinterwordspacing}{\spaceskip=\fontdimen2\font plus
\BIBentryALTinterwordstretchfactor\fontdimen3\font minus \fontdimen4\font\relax}
\providecommand{\BIBforeignlanguage}[2]{{%
\expandafter\ifx\csname l@#1\endcsname\relax
\typeout{** WARNING: IEEEtran.bst: No hyphenation pattern has been}%
\typeout{** loaded for the language `#1'. Using the pattern for}%
\typeout{** the default language instead.}%
\else
\language=\csname l@#1\endcsname
\fi
#2}}
\providecommand{\BIBdecl}{\relax}
\BIBdecl

\bibitem{pr12020270}
M.~R. Khan, Z.~M. Haider, F.~H. Malik, F.~M. Almasoudi, K.~S.~S. Alatawi, and M.~S. Bhutta, ``A comprehensive review of microgrid energy management strategies considering electric vehicles, energy storage systems, and ai techniques,'' \emph{Processes}, vol.~12, no.~2, 2024.

\bibitem{goyal2024integrating}
S.~Goyal, A.~S. Rajawat, R.~Mittal, and D.~P. Shrivastava, ``Integrating ai-enabled post-quantum models in quantum cyber-physical systems opportunities and challenges,'' \emph{Applied Data Science and Smart Systems}, pp. 491--498, 2024.

\bibitem{10734228}
C.~Ren, Z.~Y. Dong, H.~Yu, M.~Xu, Z.~Xiong, and D.~Niyato, ``Esqfl: Digital twin-driven explainable and secured quantum federated learning for voltage stability assessment in smart grids,'' \emph{IEEE Journal of Selected Topics in Signal Processing}, vol.~18, no.~5, pp. 964--978, 2024.

\bibitem{9843455}
D.~Rosch-Grace and J.~Straub, ``Analysis of the potential benefits from using quantum computing for aerospace applications,'' in \emph{2022 IEEE Aerospace Conference (AERO)}, 2022, pp. 1--6.

\bibitem{engproc2025087068}
N.~Yakubova, K.~Usmanov, Z.~Turakulov, and J.~Eshbobaev, ``Application of quantum computing algorithms in the synthesis of control systems for dynamic objects,'' \emph{Engineering Proceedings}, vol.~87, no.~1, 2025.

\bibitem{lemo2025modelisation}
A.~O. Lemo~Ngounou, ``Mod{\'e}lisation et simulations num{\'e}riques-quantiques supportant la conception d’isolation gazeuse environnementalement neutre dans les disjoncteurs de moyenne tension,'' Ph.D. dissertation, Universit{\'e} du Qu{\'e}bec {\`a} Trois-Rivi{\`e}res, 2025.

\bibitem{1571112244}
A.~Lemo, M.~Thiam, K.~I. Pierre, C.~Cossette, and A.~W. Skorek, ``Towards a digital twin of medium-voltage circuit breakers using quantum algorithms and deep learning,'' in \emph{2025 IEEE Canadian Conference on Electrical and Computer Engineering (CCECE)}, 2025, pp. 1--5.

\bibitem{10549913}
G.~Magesh \emph{et~al.}, ``Quantum channel optimization: Integrating quantum-inspired machine learning with genetic adaptive strategies,'' \emph{IEEE Access}, vol.~12, pp. 80\,397--80\,417, 2024.

\bibitem{mathur2025federatedlearningmeetsquantum}
A.~Mathur, A.~Gupta, and S.~K. Das, ``When federated learning meets quantum computing: Survey and research opportunities,'' 2025.

\bibitem{demaio2025roadhybridquantumprograms}
V.~D. Maio, I.~Brandic, E.~Deelman, and J.~Cito, ``The road to hybrid quantum programs: Characterizing the evolution from classical to hybrid quantum software,'' 2025.

\bibitem{abdullah2024uncertaintysupplychaindigital}
A.~Abdullah, F.~R. Sandjaja, A.~A. Majeed, G.~Wickremasinghe, K.~Rafferty, and V.~Sharma, ``Uncertainty in supply chain digital twins: A quantum-classical hybrid approach,'' 2024.

\bibitem{10969886}
S.~M~R, L.~G, M.~B~A, and B.~G~N, ``Addressing cybersecurity challenges in 6g networks through ai-driven adaptive defense mechanisms and quantum-resilient protocols,'' in \emph{2025 International Conference on Computing for Sustainability and Intelligent Future (COMP-SIF)}, 2025, pp. 1--12.

\bibitem{10933564}
B.~Bera, A.~K. Das, and B.~Sikdar, ``Quantum-resistant secure communication protocol for digital twin-enabled context-aware iot-based healthcare applications,'' \emph{IEEE Transactions on Network Science and Engineering}, vol.~12, no.~4, pp. 2722--2738, 2025.

\bibitem{bishwas2024strategicroadmapquantumresistant}
A.~K. Bishwas and M.~Sen, ``Strategic roadmap for quantum- resistant security: A framework for preparing industries for the quantum threat,'' 2024.

\bibitem{herian2025much}
R.~Herian, ``As much as faith: A speculation on quantum computing with fiduciary law in public governance,'' in \emph{Public Governance and Emerging Technologies: Values, Trust, and Regulatory Compliance}.\hskip 1em plus 0.5em minus 0.4em\relax Springer Nature Switzerland Cham, 2025, pp. 217--238.

\bibitem{10391714}
S.~Suhas and S.~Divya, ``Quantum-improved weather forecasting: Integrating quantum machine learning for precise prediction and disaster mitigation,'' in \emph{2023 International Conference on Quantum Technologies, Communications, Computing, Hardware and Embedded Systems Security (iQ-CCHESS)}, 2023, pp. 1--7.

\bibitem{Abreu_2024}
D.~Abreu, C.~E. Rothenberg, and A.~Abelém, ``Qml-ids: Quantum machine learning intrusion detection system,'' in \emph{2024 IEEE Symposium on Computers and Communications (ISCC)}.\hskip 1em plus 0.5em minus 0.4em\relax IEEE, Jun. 2024, p. 1–6.

\bibitem{ozguler2025performanceevaluationvariationalquantum}
A.~B. Özgüler, ``Performance evaluation of variational quantum eigensolver and quantum dynamics algorithms on the advection-diffusion equation,'' 2025.

\bibitem{SAINI202537}
K.~Saini, A.~Singh, A.~Ahuja, N.~Arora, and R.~Saini, ``Chapter 2 - research advancements in quantum computing digital twins,'' in \emph{Digital Twins for Smart Cities and Villages}, S.~Iyer, A.~Nayyar, A.~Paul, and M.~Naved, Eds.\hskip 1em plus 0.5em minus 0.4em\relax Elsevier, 2025, pp. 37--53.

\bibitem{yakubova2025application}
N.~Yakubova, K.~Usmanov, Z.~Turakulov, and J.~Eshbobaev, ``Application of quantum computing algorithms in the synthesis of control systems for dynamic objects,'' \emph{Engineering Proceedings}, vol.~87, no.~1, p.~68, 2025.

\bibitem{ali2023quantum}
M.~Z. Ali, A.~Abohmra, M.~Usman, A.~Zahid, H.~Heidari, M.~A. Imran, and Q.~H. Abbasi, ``Quantum for 6g communication: A perspective,'' \emph{IET Quantum Communication}, vol.~4, no.~3, pp. 112--124, 2023.

\bibitem{iraola2025hp2c}
E.~Iraola, M.~Garc{\'\i}a-Lorenzo, F.~Lordan-Gomis, F.~Rossi, E.~Prieto-Araujo, and R.~Badia, ``Hp2c-dt: High-precision high-performance computer-enabled digital twin,'' \emph{arXiv preprint arXiv:2506.10523}, 2025.

\bibitem{kasztelnik2023digital}
M.~Kasztelnik, P.~Nowakowski, J.~Meizner, M.~Malawski, A.~Nowak, K.~Gadek, K.~Zajac, A.~A.~L. Mattina, and M.~Bubak, ``Digital twin simulation development and execution on hpc infrastructures,'' in \emph{International Conference on Computational Science}.\hskip 1em plus 0.5em minus 0.4em\relax Springer, 2023, pp. 18--32.

\bibitem{10498081}
R.~Rayhana, L.~Bai, G.~Xiao, M.~Liao, and Z.~Liu, ``Digital twin models: Functions, challenges, and industry applications,'' \emph{IEEE Journal of Radio Frequency Identification}, vol.~8, pp. 282--321, 2024.

\bibitem{ullah2022quantum}
M.~H. Ullah, R.~Eskandarpour, H.~Zheng, and A.~Khodaei, ``Quantum computing for smart grid applications,'' \emph{IET Generation, Transmission \& Distribution}, vol.~16, no.~21, pp. 4239--4257, 2022.

\bibitem{jameil2025quantum}
A.~K. Jameil and H.~Al-Raweshidy, ``Quantum-enhanced digital twin iot for efficient healthcare task offloading,'' \emph{Discover Applied Sciences}, vol.~7, no.~6, p. 525, 2025.

\bibitem{hammadia2025quantum}
T.~Hammadia, A.~M. Saber, and D.~Kundur, ``Quantum variational circuits for detection of false data injection against power transformers,'' in \emph{2025 IEEE Industry Applications Society Annual Meeting (IAS)}.\hskip 1em plus 0.5em minus 0.4em\relax IEEE, 2025, pp. 1--8.

\bibitem{10926904}
M.~Bokhtiar Al~Zami, S.~Shaon, V.~Khanh~Quy, and D.~C. Nguyen, ``Digital twin in industries: A comprehensive survey,'' \emph{IEEE Access}, vol.~13, pp. 47\,291--47\,336, 2025.

\end{thebibliography}

\end{document}